\documentstyle[12pt,epsfig]{article}
\def\beq{\begin{equation}}
\def\eeq{\end{equation}}
\def\bea{\begin{eqnarray}}
\def\eea{\end{eqnarray}}
\def\non{\nonumber}
\def\bib{\bibitem}

\def\v{\vert}

\def\r{\rangle}

\begin{document}

\begin{center}
{\large \bf \sf Multi-band structure of a coupling constant for quantum bound 
states \\ of a generalized nonlinear Schr\"odinger model}

\vspace{1.3cm}

{\sf B. Basu-Mallick$^1$\footnote{e-mail address: biru@theory.saha.ernet.in},
Tanaya Bhattacharyya$^1$\footnote{e-mail address: tanaya@theory.saha.ernet.in}
and Diptiman Sen$^2$\footnote{e-mail address: diptiman@cts.iisc.ernet.in}}

\bigskip

{\em $^1$Theory Group, Saha Institute of Nuclear Physics, \\
1/AF Bidhan Nagar, Kolkata 700 064, India} 

\bigskip

{\em $^2$Centre for High Energy Physics, Indian Institute of Science, \\
Bangalore 560012, India}
\end{center}

\bigskip
\bigskip

\noindent {\bf Abstract}

Using Bethe ansatz we study $N$-body bound states of a generalized
nonlinear Schr\"odinger model with two coupling constants $c$ and $\eta$. 
We find that bound states exist for all values of $c$ but only within certain
ranges (called bands) of $\eta$. These ranges are governed by Farey sequences
in number theory. 

\bigskip


\vspace {.1 cm}

\newpage

\noindent \section{Introduction}
\renewcommand{\theequation}{1.{\arabic{equation}}}
\setcounter{equation}{0}

\medskip

Bound states in integrable quantum field theory models in 1+1 dimensions have
been studied extensively for many years 
\cite{mcg,thac,fadd,skyl1,skyl2,bhat,shni,kund,basu1,basu2,basu3,basu4}. The 
quantum bound states are usually constructed by using either the coordinate 
Bethe ansatz or the algebraic Bethe ansatz. For an integrable non-relativistic 
Hamiltonian, the coordinate Bethe ansatz can yield the exact eigenfunctions in
the coordinate representation. If such an eigenfunction decays sufficiently 
fast when any of the particle coordinates tends towards infinity (keeping the 
center of mass coordinate fixed), we call such a localized square-integrable 
eigenfunction a quantum bound state. It is also possible to construct quantum
bound states using the algebraic Bethe ansatz, by choosing 
appropriate distributions of the spectral parameter in the complex plane 
\cite{fadd,skyl1,skyl2}. It is usually found that localized quantum bound 
states of various integrable models, including the well known 
nonlinear Schr\"odinger model (NLS) and the sine-Gordon model, 
have positive binding energy \cite{mcg,thac,fadd,skyl1,skyl2,bhat}. 

In this paper, we will study the quantum bound 
states of a generalized NLS model. Classical and quantum versions of 
this type of models have found applications in different areas 
of physics like circularly polarized nonlinear Alfven waves in a 
plasma \cite{wada,clar} and quantum properties of solitons 
in optical fibers \cite{koda}. The Hamiltonian of the 
generalized NLS model in its second quantized form is given by \cite{shni}
\bea
H = \int_{-\infty}^{+\infty} dx ~\Big[ ~\hbar \psi_x^{\dagger} \psi_x 
+ c {\psi^\dagger}^2 {\psi}^2 +
i\eta \{{\psi^\dagger}^2 \psi \psi_x - \psi^\dagger_x {\psi^\dagger} \psi^2 
\} ~\Big] ~,
\label{a1}
\eea
where the subscripts $t$ and $x$ denote partial derivatives with respect to 
time and space respectively, $\eta$ and $c$ are real coupling 
constants, and we have set the particle mass $m = 1/2$. 
The coupling constant $\eta$ is dimensionless whereas $c$ has the 
dimension of inverse length. The field operators $ \psi(x,t),~
\psi^\dagger(x,t)$ obey the equal time commutation relations,
$[\psi(x,t), \psi(y,t)] = [\psi^\dagger(x,t), \psi^\dagger(y,t)] = 0$, and 
$[\psi(x,t), \psi^\dagger(y,t)]$ $= \hbar \delta (x-y)$. 

It may be observed that for two special cases corresponding to $\eta=0$ and 
$c=0$, the Hamiltonian of the generalized NLS model (\ref{a1}) reduces to
that of the NLS model and the derivative NLS (DNLS) model respectively. 
The quantum bound states for these integrable NLS and DNLS models have been 
investigated earlier by using both algebraic Bethe ansatz and coordinate Bethe
ansatz \cite{mcg,thac,fadd,skyl1,shni,kund,basu1,basu2,basu3,basu4}.
In this paper, our aim is to investigate the ranges of values of the coupling 
constants for which localized quantum $N$-body bound states exist for the 
generalized NLS model. In this context it should be noted that,
unlike the cases of the NLS and DNLS models,
the integrability of the generalized NLS model (\ref{a1}) has not been
established so far. So the method of algebraic Bethe ansatz can not be used 
at present to study this model. 

It may also be mentioned that an analysis of the classical NLS model suggests 
a fascinating application of the spectral analysis for nonlinear operators, 
which has been studied earlier \cite{rabin}. However, for the generalized NLS 
model, we shall take the following approach. Rather than trying to deal 
directly with the spectral analysis of the model defined in (\ref{a1}), we 
shall project the second quantized Hamiltonian to the bosonic N-particle 
subspace and reduce it to a linear operator. The method of coordinate Bethe 
ansatz can then be used to find the spectrum of such a linear operator.
In Sec. 2, we apply this coordinate Bethe ansatz to find out the conditions 
which the Bethe momenta have to satisfy in order that a quantum $N$-body bound 
state should exist. In Sec. 3, we analyze these conditions on Bethe momenta and
find that such bound states exist for all possible values (both positive and 
negative) of $c$ and within several non-overlapping ranges (called bands) of 
$\eta$. We also apply the idea of Farey sequences in number theory to 
completely determine the ranges of all bands for which $N$-body bound states 
exist for a given value of $N$. In Sec. 4, we show that the bound states 
appearing within each band can have both positive and negative momentum. 
Moreover, bound states with zero momentum can be constructed for a wide range 
of the coupling constants. We also calculate the binding energy 
of the bound states and find that it can take both positive and negative 
value within each band. In Sec. 5 we make some concluding remarks.

\vspace{1cm}

\noindent \section{Conditions for quantum $N$-body bound states in the
generalized NLS model}
\renewcommand{\theequation}{2.{\arabic{equation}}}
\setcounter{equation}{0}

\medskip

To apply the coordinate Bethe ansatz, we separate the full bosonic Fock space 
associated with the Hamiltonian (\ref{a1}) into disjoint $N$-particle subspaces
$\vert S_N \rangle $. We want to solve the eigenvalue equation $H \v S_N \r = E
\v S_N \r $. The coordinate representation of this equation is given by
\bea
H_N ~\tau_N (x_1, x_2, \cdots , x_N ) ~=~ E ~\tau_N (x_1, x_2, \cdots , x_N )~,
\label{b1}
\eea
where the $N$-particle symmetric wave function $\tau_N( x_1, x_2,\cdots , 
x_N)$ is defined as 
\bea
\tau_N( x_1, x_2, \cdots , x_N ) ~=~ \frac{1}{\sqrt{N!}} ~\langle 0 \vert 
\psi(x_1) \cdots \psi(x_N) \vert S_N \rangle ~,
\label{b2}
\eea 
and $H_N$, the projection of the second-quantized Hamiltonian $H$ 
(\ref{a1}) on to the $N$ particle sector, is given by
\beq
H_N ~=~ -\hbar^2 ~\sum_{j=1}^N ~\frac{\partial^2}{\partial x_j^2} ~+~ 2\hbar^2
c~\sum_{l<m}~\delta(x_l - x_m)~+~2i \hbar^2 \eta ~\sum_{l<m} ~\delta (x_l - 
x_m )~ \Big( \frac{\partial}{\partial x_l} + \frac{\partial} {\partial x_m} 
\Big).
\label{b3}
\eeq

It is evident that $H_N$ commutes with the total momentum operator in 
the $N$-particle sector, which is defined as
\beq
P_N ~=~ -i\hbar ~\sum_{j=1}^N ~\frac{\partial}{\partial x_j} ~. 
\label{b4}
\eeq

Note that $H_N$ remains invariant while $P_N$ changes sign if we change the 
sign of $\eta$ and transform all the $x_i \rightarrow - x_i$, keeping $c$ 
unchanged; let us call this the parity transformation. Hence it is sufficient 
to study the model for $\eta >0$. The eigenfunctions for $\eta < 0$ can then 
by obtained by changing $x_i \rightarrow -x_i$; this leaves the energy 
invariant but reverses the momentum.

Let us first construct the eigenfunctions 
of the Hamiltonian (\ref{b3}) for the two-particle case, without imposing any
symmetry property on $\tau_2(x_1,x_2)$ under the exchange of the particle 
coordinates. For the region $x_1<x_2$, we may take the eigenfunction to be
\bea
\tau_2 (x_1, x_2) ~=~ \exp ~\{ i(k_1 x_1 + k_2 x_2) \} ~, 
\label{b5}
\eea
where $k_1$ and $k_2$ are two distinct wave numbers. 
Using Eq. (\ref{b1}) for $N=2$, we find that this two-particle 
wave function takes the following form in the region $x_1>x_2 ~$:
\bea
\tau_2(x_1,x_2) ~=~ A(k_1,k_2)\exp ~\{ i(k_1 x_1 + k_2 x_2) \} + B(k_1,k_2)
\exp ~\{ i(k_2 x_1 + k_1 x_2) \} ~,
\label{b6}
\eea
where the `matching coefficients' $A(k_1,k_2)$ and $B(k_1,k_2)$ are given by 
\cite{shni}
\beq
A(k_1,k_2) ~=~ \frac{k_1 - k_2 + i \eta (k_1+ k_2) - ic}{k_1 - k_2} ~, ~~~~
B(k_1,k_2) ~=~ 1- A(k_1 , k_2) ~. 
\label{b7}
\eeq
By using these matching coefficients, we can construct completely symmetric
$N$-particle eigenfunctions for the Hamiltonian (\ref{b3}). In the region
$x_1< x_2 < \cdots < x_N$, these eigenfunctions are given by \cite{shni,gutk}
\beq
\tau_N (x_1, x_2 , \cdots , x_N) ~=~ \sum_\omega \left (\prod_{l<m}
\frac{A(k_{\omega(m)},k_{\omega(l)})}{A(k_m,k_l)}\right) \rho_{\omega(1), 
\omega(2), \cdots , \omega(N)} (x_1, x_2, \cdots , x_N) ~,
\label{b8}
\eeq
where 
\beq
\rho_{\omega(1), \omega(2), \cdots , \omega(N)} (x_1, x_2, \cdots , x_N) ~=~
\exp ~\{ i (k_{\omega(1)}x_1 + \cdots + k_{\omega(N)} x_N ) \} ~.
\label{b9}
\eeq
In the expression (\ref{b8}), the $k_n$'s are all distinct wave numbers, 
and $\sum_{\omega}$ implies summing over all
permutations of the integers $(1,2,....N)$. The eigenvalues of 
the momentum (\ref{b4}) and Hamiltonian (\ref{b3}) operators, corresponding 
to the eigenfunctions $\tau_N(x_1, x_2, \cdots , x_N)$, are given by
\bea
&&~~~~~~~~P_N ~\tau_N(x_1, x_2, \cdots , x_N) ~=~ \hbar 
\Big(\sum_{j=1}^N k_j \Big) ~\tau_N(x_1, x_2, \cdots , x_N) ~, \non
~~~~~~~~~~~~~~~~~~~~ (2.10a) \\
&&~~~~~~~~H_N ~\tau_N(x_1, x_2, \cdots , x_N) ~=~ 
\hbar^2 \Big(\sum_{j=1}^N k_j^2 \Big) ~\tau_N(x_1, x_2, \cdots , x_N) ~. 
\non ~~~~~~~~~~~~~~~~~~~ (2.10b)
\eea
\addtocounter{equation}{1}

The wave function in (\ref{b8}) will represent a localized bound state if 
it decays when any of the relative coordinates measuring the distance between
a pair of particles tends towards infinity. To obtain the condition for the 
existence of such a localized bound state, let us consider the following
wave function in the region $x_1<x_2<\cdots <x_N$:
\bea
\rho_{1,2, \cdots ,N} (~ x_1,x_2, \cdots , x_N ~) ~=~ \exp ~(i\sum_{j=1}^N 
k_j x_j) ~.
\label{b11}
\eea
As before, the momentum eigenvalue corresponding to this wave function is 
given by $\hbar \sum_{j=1}^N k_j$. Since this must be a real quantity, we 
obtain the condition
\beq
\sum_{j=1}^N q_j ~=~ 0 ~,
\label{b12}
\eeq
where $q_j$ denotes the imaginary part of $k_j$. The probability density 
corresponding to the wave function $\rho_{1,2, \cdots ,N} (~ x_1,x_2, \cdots ,
x_N ~)$ in (\ref{b11}) can be expressed as 
\bea
{|\rho_{1,2,\cdots ,N} (~ x_1,x_2, \cdots , x_N ~)|}^2 ~=~ \exp ~\Big\{ ~
2 \sum_{r=1}^{N-1} \Big(\sum_{j=1}^r q_j\Big) ~y_r ~\Big \} ~,
\label{b13}
\eea
where the $y_r$'s are the $N-1$ relative coordinates: $y_r \equiv 
x_{r+1} - x_r$, and we have used Eq. (\ref{b12}). It is evident that the 
probability density in (\ref{b13}) decays exponentially in the limit 
$y_r \rightarrow \infty$ for one or more values of $r$, provided that all 
the following conditions are satisfied:
\beq
q_1< 0 ~, ~~~~q_1+q_2 < 0 ~, ~~\cdots\cdots ~~, ~\sum_{j=1}^{N-1} ~q_j < 0 ~. 
\label{b14}
\eeq
Equations (\ref{b12}) and (\ref{b14}) imply that the wave function 
$\rho_{1,2,\cdots,N}(x_1, x_2, \cdots, x_N)$ (\ref{b11}) is square-integrable 
if one holds the centre-of-mass coordinate $X={\sum_i x_i}/ N$ fixed, and 
integrates over the relative coordinates $y_r$. In the region $x_1<x_2< \cdots
<x_N$, the integrals over the $y_r$'s all run from 0 to $\infty$, and they are
independent of each other. The probability density (\ref{b13}) is independent 
of $X$ and due to the conditions (\ref{b14}), the integration of this 
probability density over the $y_r$'s gives a finite result.

Note that the wave function (\ref{b11}) is obtained by taking $\omega$ as the 
identity permutation in (\ref{b9}). However, the full wave function (\ref{b8}) 
also contains terms like (\ref{b9}) with $\omega$ representing all possible 
nontrivial permutations. The conditions which ensure the decay of such a term
with a nontrivial permutation will, in general, contradict the 
conditions (\ref{b14}). To construct a decaying wave function, therefore, 
the coefficients of all terms $\rho_{\omega(1), \omega(2), \cdots , \omega(N)}
(x_1, x_2, \cdots , x_N)$ with nontrivial permutations 
must be made to vanish. It turns out that it is sufficient to require:
\beq
A( k_{1}, k_{2} ) ~=~ 0, ~~A( k_{2}, k_{3} ) ~=~ 0, ~\cdots\cdots ~,~~ 
A( k_{N-1}, k_{N}) ~=~ 0 ~.
\label{b15}
\eeq

Thus the simultaneous validity of the conditions
(\ref{b12}), (\ref{b14}) and (\ref{b15}) ensures that the full wave function
$\tau_N(x_1, x_2, \cdots , x_N)$ (\ref{b8}) represents a localized
bound state. Using the conditions (\ref{b12}) and (\ref{b15}), one can 
obtain an expression for the complex $k_n$'s in the form 
\beq
k_n ~=~ \chi ~e^{-i(N+1-2n)\phi}+\frac{c}{2 \tan \phi} ~,
\label{b16}
\eeq
where $\chi$ is a real parameter, and $\phi$ is related to the coupling 
constant as 
\beq
\phi ~=~ \tan^{-1} (\eta ) ~.
\label{b17}
\eeq
To obtain an unique value of $\phi$ from the above equation, we restrict it
to the fundamental region $-\frac{\pi}{2} < \phi (\neq 0) < \frac{\pi}{2}$. 
[Note that $\eta$ and $\phi$ have the same sign.
Due to the parity symmetry mentioned above, we can restrict our 
attention to the range $0 < \phi < \frac{\pi}{2}$].

Now, let us verify whether the $k_n$'s in (\ref{b16}) satisfy the conditions 
(\ref{b14}). Summing over the imaginary parts of these $k_n$'s, we can express 
the conditions (\ref{b14}) in the form 
\beq
\chi ~\frac{\sin (l \phi)}{\sin \phi} ~\sin [(N-l) \phi] ~>~ 0 \quad {\rm for}
\quad l ~=~ 1, ~2, ~\cdots ~, N-1 ~. 
\label{b18}
\eeq
Thus, for some given values of $c, \phi$, $N$ and $\chi$, a bound state will 
exist when all the above inequalities are simultaneously satisfied. 
Surprisingly, the above condition for having a quantum bound state for the 
generalized NLS model is completely independent of the coupling constant $c$.
By using Eqs. (\ref{b16}) and (2.10a), we obtain the momentum eigenvalue of 
such a bound state to be
\beq
P ~=~ \frac{\hbar c N}{2 \tan \phi} + \hbar \chi ~\frac{\sin (N\phi)}
{\sin \phi}\, .
\label{b19}
\eeq
Hence $\chi$ is related to the momentum as
\beq
\chi ~=~ \frac{1}{\hbar}(P-\frac{\hbar c N}{2 \tan \phi})\frac{\sin \phi}
{\sin (N \phi)}\, .
\label{b19a}
\eeq

In a similar way, the energy eigenvalue of these bound states is found to be 
\beq
E ~=~ \frac{\hbar^2 \chi^2 \sin(2N \phi)}{\sin(2\phi)} ~+ \frac{\hbar^2 c^2 N}
{4 \tan^2 \phi} + \frac{\hbar^2\chi c \sin (N \phi) \cos \phi}{\sin ^2 \phi}.
\label{b20}
\eeq
For some fixed value of $c, \phi $ and $N$, the above expression for the energy
has a minimum for a nonzero value of the momentum given by $$ P_0 = 
\frac{N \hbar c}{2 \tan \phi}(1- \frac{\tan (N\phi)}{N \tan \phi})\, ,$$ 
provided that $\tan (N\phi) > 0$. The next section of our paper will be devoted
to finding the ranges of values of $\phi$ where all the inequalities 
(\ref{b18}) are simultaneously satisfied for a given value of the particle 
number $N$.

\vspace{1cm}

\noindent \section{Determining the values of $\phi$ where $N$-body bound states
exist}
\renewcommand{\theequation}{3.{\arabic{equation}}}
\setcounter{equation}{0}

\medskip

In this section, we will study the values of $\phi$ where $N$-body bound
states exist for different values of $N$ for the generalized NLS model. 
In an earlier analysis, the quantum bound states of this model had been 
found to exist for $0<\phi<\frac{\pi}{N}$ \cite{shni}. However from our 
analysis, it will turn out that there are several non-overlapping 
ranges (bands) of $\phi$ for which quantum bound states exist for the 
generalized NLS model. Even the lowest band for which such bound states 
exist (i.e., when $\phi$ lies in the range $0<\phi<\frac{\pi}{N-1}$) is
wider than that obtained earlier \cite{shni}. Since eqn. (\ref{b18}) is 
independent of the coupling constant $c$, bound states are formed within the 
above mentioned bands for all possible values of $c$. 
Consequently, in contrast to the case of NLS model where bound states appear 
only for negative values of $c$, bound states of the generalized NLS model 
can appear for both positive and negative values of $c$. 

Incidentally, it is quite remarkable that our results concerning the 
band structure of the generalized NLS model 
turn out to be very similar to our analysis on the quantum bound states of 
the DNLS model \cite{basu3,basu4}. For the simplest case $N=2$, 
the condition (\ref{b18}) is satisfied when $\phi$ lies in the range
$0 < \phi < \frac {\pi}{2}$ ($- \frac {\pi}{2} < \phi < 0$)
for the choice $\chi >0$ ($\chi < 0$). Thus any nonzero value of $\phi$
within its fundamental region can generate a $2$-body bound state. 

We will now consider the more interesting case with $N \geq 3$. Due to the 
parity symmetry of the Hamiltonian in (\ref{b3}), we will henceforth assume 
that $\phi > 0$.

Let us consider a value of $\phi$ of the form
\beq
\phi_{N,n} ~\equiv ~\frac{\pi n}{N} ~,
\label{phinn}
\eeq
where $n$ is an integer satisfying $1 \le n < N/2$. 

It can be easily shown that all the inequalities in (\ref{b18}) are satisfied 
for $\phi = \phi_{N,n}$, if and only if $N$ and $n$ are relatively prime (with
$n$ odd for $\chi >0$, and $n$ even for $\chi < 0$) \cite {basu4}. By 
continuity, it then follows that all the inequalities will hold in a 
neighborhood of $\phi_{N,n}$ extending from a value $\phi_{N,n,-}$ to a value 
$\phi_{N,n,+}$, such that $\phi_{N,n,-} < \phi_{N,n} < \phi_{N,n,+}$. We will 
call the region
\beq
\phi_{N,n,-} < \phi < \phi_{N,n,+}
\eeq
as the band $B_{N,n}$. In this band, there is a bound state with $N$ particles.

We now have to determine the end points $\phi_{N,n,-}$ and $\phi_{N,n,+}$ of 
the band $B_{N,n}$. The inequalities in (\ref{b18}) show that the end points 
are given by $\phi$ of the form 
\beq
\phi ~=~ \frac{\pi j}{l} ~,
\label{jl}
\eeq
where $j$ and $l$ are relatively prime and satisfy the conditions: 
\beq
1 ~\le ~ l ~<~ N ~, \quad {\rm and} \quad j ~< ~\frac{l}{2}.
\label{c6}
\eeq
Thus the end points of the band $B_{N,n}$ are
given by two rational numbers $\phi /\pi$ of the form $j/l$ which lie {\it 
closest} to (and on either side of) the point $\phi_{N,n} /\pi = n/N$. 
The solution to this problem is well known in
number theory and is described by the Farey sequences \cite{nive}.

For a positive integer $N$, the Farey sequence $F_N$ is defined to be the set 
of all the fractions $a/b$ in increasing order such that (i) $0 \le a \le b 
\le N$, and (ii) $a$ and $b$ are relatively prime. 
\newpage
The Farey sequences for the first few integers are given by
\bea
F_1: & & \quad \frac{0}{1} ~~~~\frac{1}{1} \non \\
F_2: & & \quad \frac{0}{1} ~~~~\frac{1}{2} ~~~~\frac{1}{1} \non \\
F_3: & & \quad \frac{0}{1} ~~~~\frac{1}{3} ~~~~\frac{1}{2} ~~~~\frac{2}{3} ~~~~
\frac{1}{1} \non \\
F_4: & & \quad \frac{0}{1} ~~~~\frac{1}{4} ~~~~\frac{1}{3} ~~~~\frac{1}{2} ~~~~
\frac{2}{3} ~~~~\frac{3}{4} ~~~~\frac{1}{1} \non \\
F_5: & & \quad \frac{0}{1} ~~~~\frac{1}{5} ~~~~\frac{1}{4} ~~~~\frac{1}{3} ~~~~
\frac{2}{5} ~~~~\frac{1}{2} ~~~~\frac{3}{5} ~~~~\frac{2}{3} ~~~~
\frac{3}{4} ~~~~\frac{4}{5} ~~~~\frac{1}{1}
\eea

To return to our problem, we now see that the points $\phi_{N,n}$ in 
(\ref{phinn}) (which lie in the bands $B_{N,n}$) have a one-to-one 
correspondence with the fractions $n/N$, which appear on the left side of 
$1/2$ within the sequence $F_N$. Due to Eqs. (\ref{jl}) and (\ref{c6}), the 
end points of the band $B_{N,n}$ are given by
\bea
\phi_{N,n,-} ~=~ \frac{\pi a_1}{b_1} ~, ~~~~
\phi_{N,n,+} ~=~ \frac{\pi a_2}{b_2} ~,
\label{c9}
\eea
where $a_1/b_1$ and $a_2/b_2$ are the fractions lying immediately to the left
and right of $n/N$ in the Farey sequence $F_N$. These are the two unique 
fractions which lie closest to (and on either side) of $n/N$. 

In Table 1, we show the ranges of values of $\phi$ for which bound states exist
for $N=2$ to 9. 

\vspace{0.4cm}
\begin{center}
\begin{tabular}{|c|c|c|} \hline
$N$ & $n$ & Range of values \\
& & of $\phi /\pi$ \\ \hline
2 & 1 & $0 < \phi /\pi < 1/2$ \\
3 & 1 & $0 < \phi /\pi < 1/2$ \\
4 & 1 & $0 < \phi /\pi < 1/3$ \\
5 & 1 & $0 < \phi /\pi < 1/4$ \\
5 & 2 & $1/3 < \phi /\pi < 1/2$ \\
6 & 1 & $0 < \phi /\pi < 1/5$ \\
7 & 1 & $0 < \phi /\pi < 1/6$ \\
7 & 2 & $1/4 < \phi /\pi < 1/3$ \\
7 & 3 & $2/5 < \phi /\pi < 1/2$ \\
8 & 1 & $0 < \phi /\pi < 1/7$ \\
8 & 3 & $1/3 < \phi /\pi < 2/5$ \\
9 & 1 & $0 < \phi /\pi < 1/8$ \\
9 & 2 & $1/5 < \phi /\pi < 1/4$ \\
~~9~~ & ~~4~~ & ~~$3/7 < \phi /\pi < 1/2$~~ \\ \hline
\end{tabular}
\end{center}
\vspace{0.2cm}

\centerline{Table 1. The range of values of $\phi /\pi$ for which bound states
exist for various values of $N$.}

\vspace{1cm}

\noindent \section{Momentum and binding energy of a $N$-body bound state}
\renewcommand{\theequation}{4.{\arabic{equation}}}
\setcounter{equation}{0}

\medskip
In the previous section, we have determined the band structure associated with 
the generalized NLS model by proceeding in the same way as in the case of the 
DNLS model. However it should be noted that, in the case of the DNLS model, one
does not get quantum bound states with zero momentum expect for some discrete 
values of the coupling constant $\phi = \pi n/N$ \cite{basu4}. Thus, in 
general, bound states of the DNLS model do not exist in the centre-of-mass 
(COM) frame. This may be due to the fact that in the Hamiltonian of the DNLS 
model there is no dimensionful coupling constant, and it is invariant under a 
scale transformation. On the other hand, the Hamiltonian of the generalized NLS
model (\ref{a1}) has a dimensionful coupling which breaks the scale invariance.
So, we may expect to get zero momentum bound states for this model within a 
wide range of the coupling constants. The binding energy of such zero momentum 
states may then be interpreted as their internal energy in the COM frame. With 
this aim in mind, in this section we will calculate the momentum and binding 
energy for the $N$-body bound states described above.

We first look at the momentum of the bound states in a particular band 
$B_{N,n}$ using Eq.(\ref{b19}). The form of the end points given in (\ref{c9})
shows that $\sin (N \phi) =0$ at only one point in the band $B_{N,n}$, namely,
at $\phi = \phi_{N,n}$. In the right part of the band (i.e., from $\phi_{N,n}$
to $\phi_{N,n,+}$), the sign of $\sin (N\phi)$ is $(-1)^n$. In the left part 
of the band (i.e., from $\phi_{N,n,-} $ to $\phi_{N,n}$), the sign of $\sin 
(N\phi)$ is $(-1)^{n+1}$. Now, the analysis given above showed that $\chi$ 
has the same sign as $(-1)^{n+1}$. Hence the momentum given in (\ref{b19}) is 
greater (less) than ${\hbar cN}/({2\tan\phi})$ in the left (right) part of a 
band, and is equal to ${\hbar cN}/({2\tan\phi})$ at $\phi = \phi_{N,n}$. 

In Figs. 1a and 1b, we have plotted the allowed values of momentum ($P$) (\ref
{b19}) as a function of $\phi/\pi$ for $N=5$. In Fig. 1a, (taking $\hbar c =1$)
the minimum (maximum) value of $P$ in the left (right) part of a band is
$5/{2 \tan\phi}$. Similarly, in Fig. 1b (taking $\hbar c =-1$) the minimum 
(maximum) value of $P$ in the left (right) part of a band is $-5/{2 \tan\phi}$.

Next, we look at the energy using Eq. (\ref{b20}). 
To calculate the binding energy, we consider a reference state in which the 
momentum $P$ of the $N$-body bound state given in (\ref{b19}) is equally 
distributed amongst $N$ single-particle scattering states. The real 
wave number associated with each of these single-particle states is denoted 
by $k_0$. From Eqs. (2.10a) and (\ref{b19}), we obtain 
\beq
k_0 ~=~ \frac{c}{2\tan \phi} + \frac{\chi \sin (N \phi)}{N \sin \phi}. 
\label{d2}
\eeq
Using Eq. (2.10b), we can calculate the total energy for the $N$ 
single-particle scattering state as
\bea
E_s ~=~ \hbar^2 N k_0^2 ~=~ \frac{\hbar^2 \chi^2 \sin^2 (N \phi)}{N\sin^2 
\phi} ~+ \frac{\hbar^2 c^2 N}{4 \tan^2 \phi} + \frac{\hbar^2 \chi c 
\sin (N\phi) \cos \phi} {\sin ^2 \phi}.
\label{d3}
\eea
Subtracting $E$ in (\ref{b20}) from $E_s$ in (\ref{d3}), we obtain the binding 
energy of the $N$-body bound state as 
\beq
E_B (\phi, N) \equiv E_s - E ~=~ \frac{\hbar^2 \chi^2\sin (N \phi)}{\sin \phi}
\Big\{\frac{\sin (N \phi)}{N\sin \phi} -\frac{\cos (N \phi)}{\cos \phi} 
\Big\} ~.
\label{d4}
\eeq
It may be noted that the above expression for the binding energy remains 
invariant under the transformation $\phi \rightarrow -\phi$. 
Now, it can be easily shown that this expression for the binding energy
is positive in the left part of each band, negative in the right part, and 
zero at the point $\phi=\phi_{N,n}$ \cite{basu4}.

We thus see that for $\phi >0$, $(P-{\hbar c N}/{2\tan \phi})$ and the binding 
energy are both positive in the left part of a band, and they are both negative
in the right part of a band. Consequently, for the case $\phi >0$ and $c>0~ 
(c<0)$, bound states with zero momentum can be constructed within the right 
(left) side of each band. Using eqn. (\ref{b18}) it is easy to see that, the 
momentum of a quantum bound state vanishes for 
\beq 
\chi = -{cN \over 2} {\cos \phi \over \sin (N\phi)}.
\label{d5}
\eeq
Substituting this value of $\chi$ to eqn. (\ref{d4}), we obtain 
the binding energy for these zero momentum bound states as
\beq
E_B^{P=0} (\phi, N)= \frac{\hbar^2 c^2 N}{4 \tan^2 \phi }
\Big[ 1 - \frac{N \tan\phi}{\tan (N\phi)}\Big]~\, .
\label{d6}
\eeq
This can be interpreted as the internal energy of a quantum bound state in 
its COM frame.

In Fig. 2a and 2b, we have plotted the above expression for the binding energy
(\ref{d6}) as a function of $\phi/\pi$ for $N=5$ ($\hbar c =1$ in Fig. 2a and
$\hbar c=-1$ in Fig. 2b).

Finally, we want to discuss a subtle connection between quantum bound states of
the generalized NLS model and the ordinary NLS model. It has been already 
mentioned that, in the limit $\phi \rightarrow 0$, the ordinary NLS model can 
be obtained from the generalized NLS model. It is well known that quantum bound
states of NLS model exist only for $c<0$ with the $N$ momenta being given by
\beq
k_n ~=~ \frac{P}{N\hbar} ~+~ i \frac{c}{2} ~(N+1-2n) ~,
\label{d7}
\eeq
and the binding energy for such state taking the form 
\beq
 E_B(N)=\hbar^2 c^2 N (N^2 -1)/12.
\label{d8}
\eeq
On the other hand, in this article we have seen that, for any small value of 
the parameter $\phi$ lying within the lowest band, quantum bound states of the 
generalized NLS model exist for both $c>0$ and $c<0$. So it is natural to ask 
why bound states corresponding to $c>0$ disappear in the limit $\phi 
\rightarrow 0$. To answer this question, we first try to take the
$\phi \rightarrow 0$ limit of the complex momenta $k_n$ (\ref{b16})
associated with the generalized NLS model and check whether they 
reduce to the $k_n$ given in eqn. (\ref{d7}). However it is evident that, 
all $k_n$ in eqn. (\ref{b16}) as well as the corresponding total momentum $P$
(\ref{b19}) will diverge if we take the limit $\phi \rightarrow 0$ by keeping
$\chi$ fixed. To bypass this problem, we have to take the limit $\phi 
\rightarrow 0$ in a way such that the total momentum $P$ (\ref{b19}) remains 
constant. For the sake of convenience we choose $P=0$, for which the value 
of $\chi$ is given in eqn. (\ref{d5}). Substituting this value 
of $\chi$ in eqn. (\ref{b16}) and expanding the right hand side 
of this equation in a power series of $\phi$, we easily obtain 
\beq
k_n ~=~ i \frac{c}{2} ~(N+1-2n) ~+~ {\cal O}(\phi),
\label{d9}
\eeq
which reproduces the value of $k_n$ in eqn. (\ref{d7}) for $P=0$.
Similarly, by taking the limit $\phi\rightarrow 0$ of the binding energy 
(\ref{d6}) corresponding to the zero momentum bound states of the generalized
NLS model, we can reproduce $E_B(N)$ in eqn. (\ref{d8}). 
In the case of the generalized NLS model with $c>0$, we have already 
observed that zero momentum bound states exist only in the right side of the 
each band. For taking the limit $\phi\rightarrow 0$, however, we have to 
move to the left side of the lowest band. Consequently, it is impossible to
take the limit $\phi\rightarrow 0$ within the lowest band after fixing $P=0$ 
(same conclusion can be drawn for any other fixed value of $P$). On the other 
hand, for $c<0$, zero momentum bound states exist in the left side of each band
and one can easily take the limit $\phi\rightarrow 0$ after fixing $P=0$. This
fact clearly explains why quantum bound states remain in the limit $\phi 
\rightarrow 0$ only if $c<0$. 

\vspace{1cm}

\noindent \section{Conclusion}

\medskip

By applying the coordinate Bethe ansatz, we have investigated the range of the 
coupling constants ($c$,$\phi$) for which localized quantum $N$-body bound 
states exist in a generalized NLS model. It is found that such bound states 
exist for all possible values (both positive and negative) of $c$ and within 
several non-overlapping ranges (called bands) of $\phi$. Using the ideas of 
Farey sequences appearing in number theory, we have given explicit expressions
for all the allowed bands of $\phi$ for which $N$-body bound states exist. 


We have also calculated the momentum and binding energy
for the bound states within all bands in the region $\phi>0$. It is found 
that bound states with positive (negative) values of $(P-{\hbar c N}/{2\tan 
\phi})$ appear in the left (right) part of each band, and these states have 
positive (negative) binding energy. Consequently, bound states
with zero momentum can be constructed within a wide range of the 
coupling constants. This happens in contrast to the case of the DNLS model, 
where zero momentum bound states appear for only some discrete values of 
$\phi$. The binding energy of zero momentum states has also been calculated 
and interpreted as the internal energy in the centre-of-mass frame. We have 
also remarked on some subtleties of taking the limit $\phi \rightarrow 0$
for the bound states of the generalized NLS model, and explained why 
this process does not lead to any bound state of NLS model for $c>0$.

It is currently not known if the generalized NLS model discussed here is 
integrable. If the model turns out to be non-integrable, then the bound states
which we have found using the Bethe ansatz may not be complete, and there could
be some other bound states. This may be an interesting problem for future 
studies.


\newpage

\begin{figure}[htb]
\vspace*{-1.2cm}
\begin{center}
\epsfig{figure=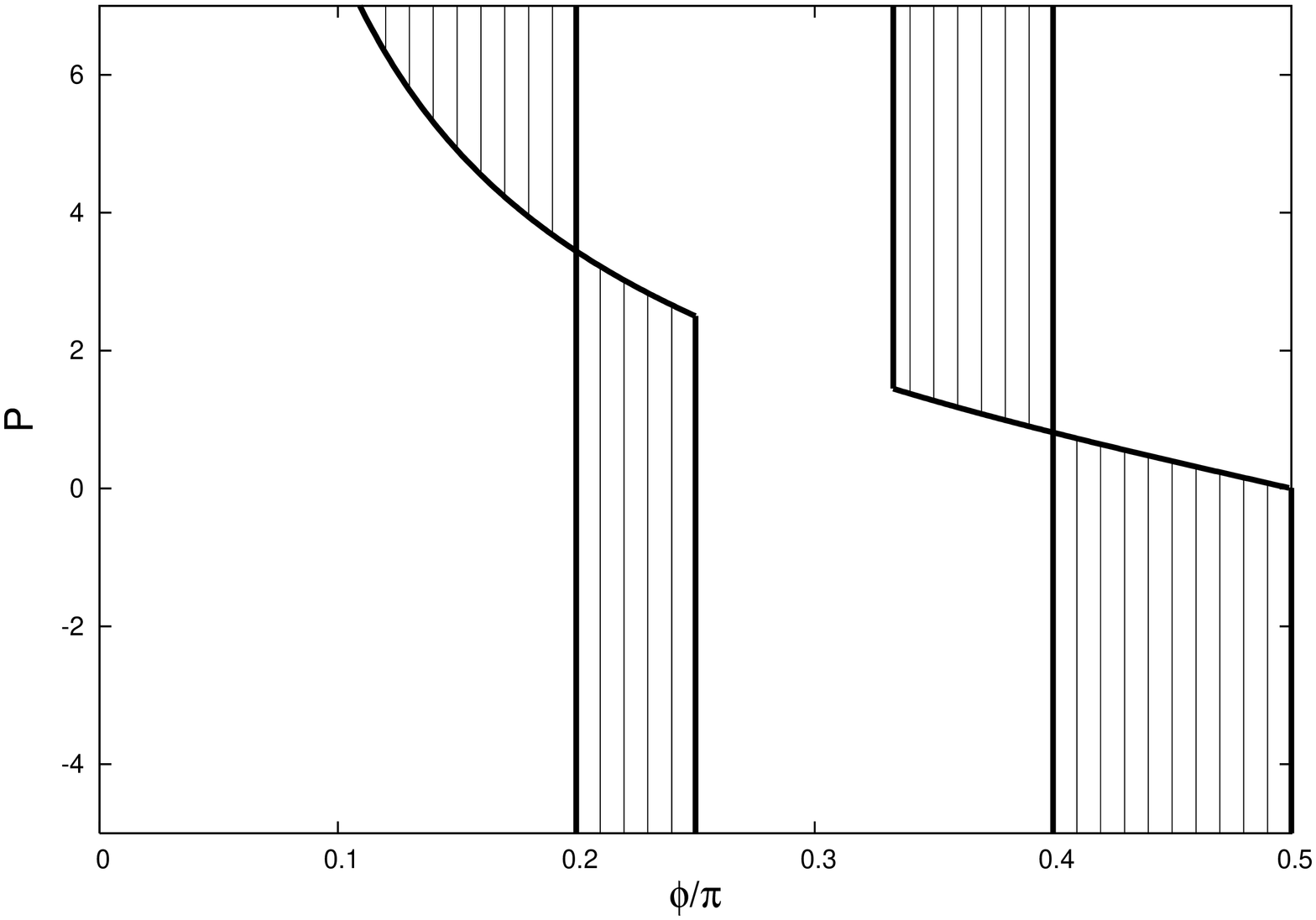,height=10cm,width=8cm,angle=-90}
\centerline{(a)}
\epsfig{figure=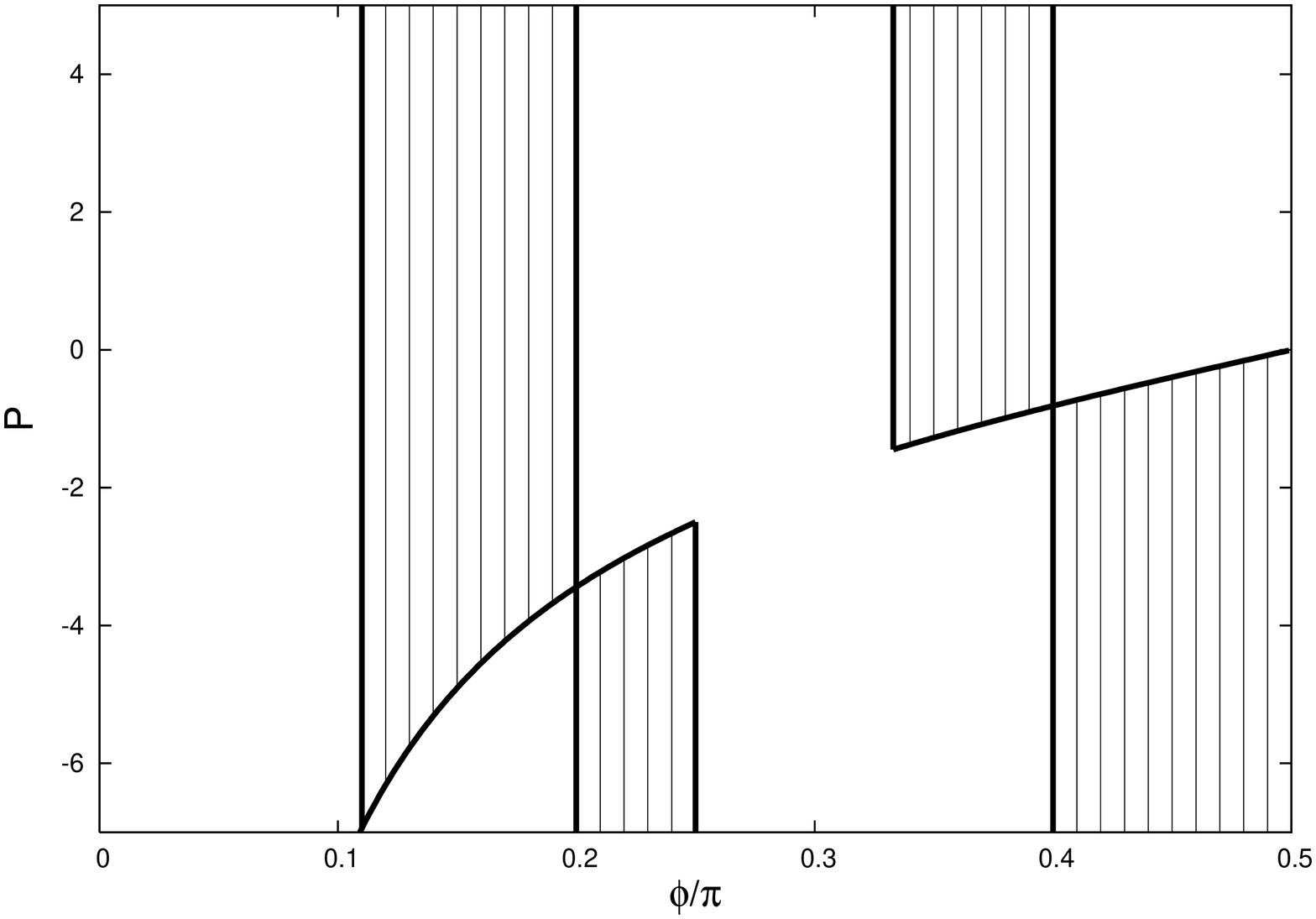,height=10cm,width=8cm,angle=-90}
\centerline{(b)}
\end{center}
\end{figure}
\noindent{Fig. 1. The filled in regions indicate the values of the momentum 
as a function of $\phi /\pi$ for which a 5-body bound state exists. In 
figures (a) and (b), $\hbar c =1$ and $-1$ respectively.}

\newpage

\begin{figure}[htb]
\vspace*{-1.2cm}
\begin{center}
\epsfig{figure=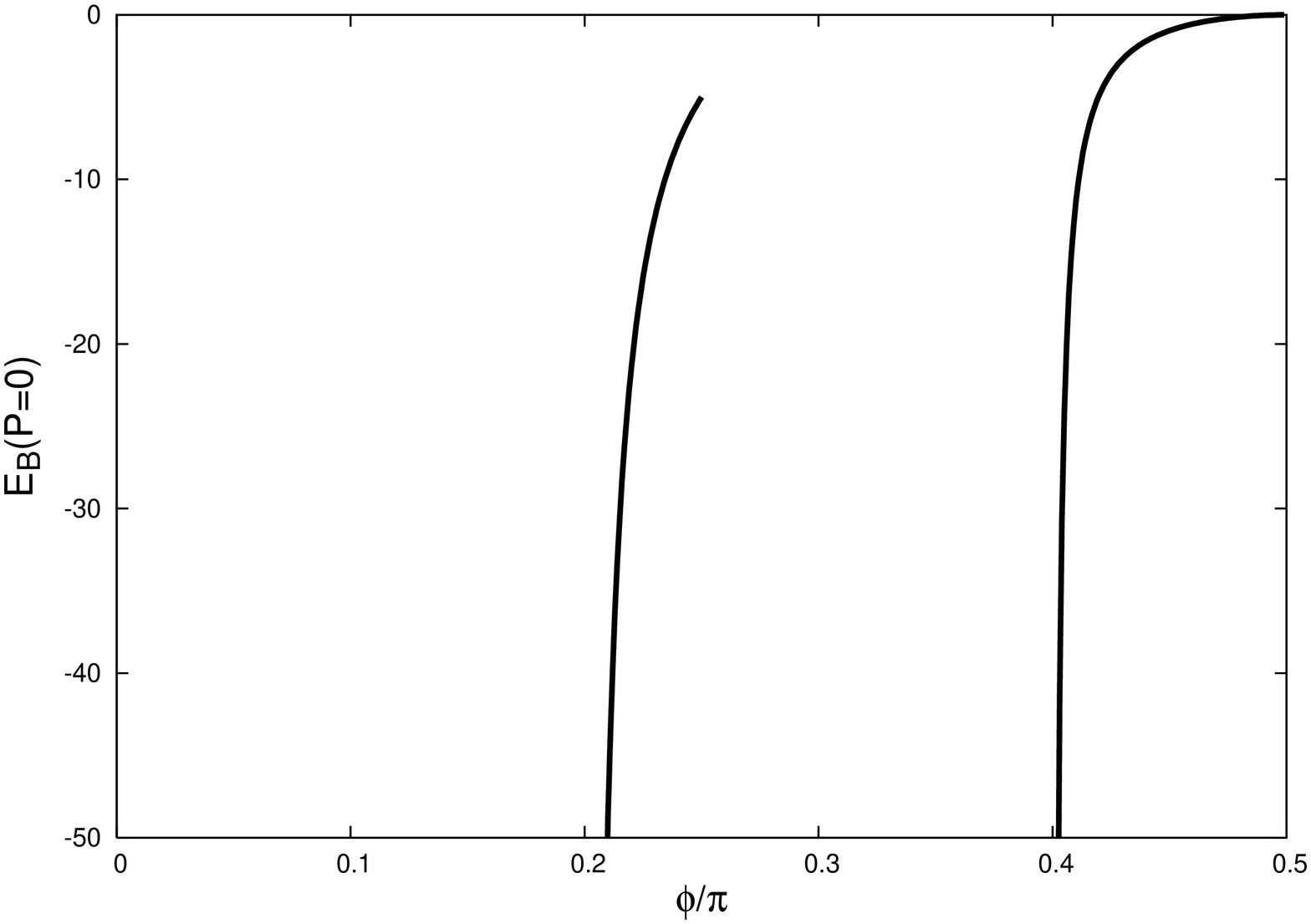,height=10cm,width=8cm,angle=-90}
\centerline{(a)}
\epsfig{figure=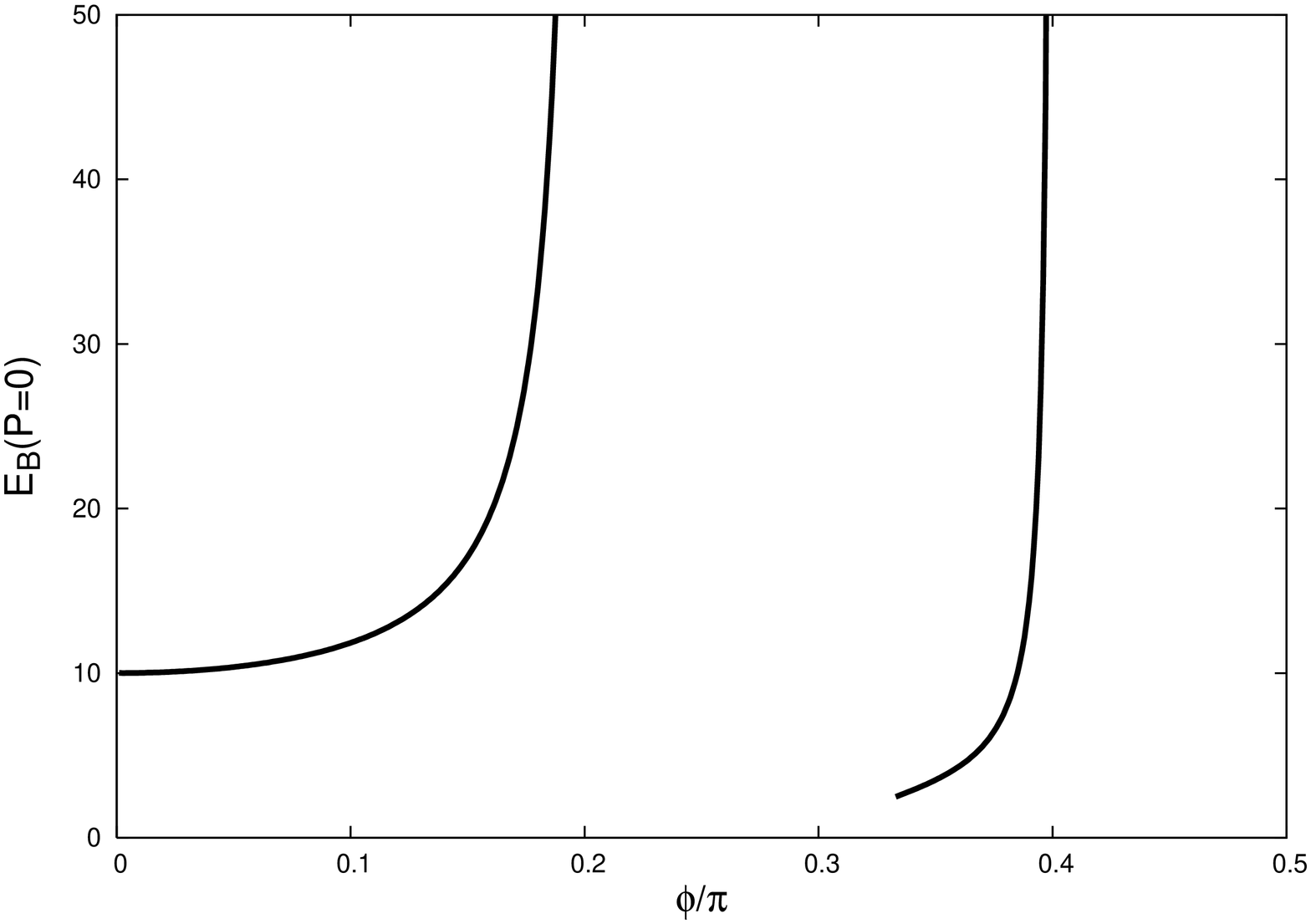,height=10cm,width=8cm,angle=-90}
\centerline{(b)}
\end{center}
\end{figure}
\noindent{Fig. 2. The binding energy of the zero momentum 5-body bound state 
as a function of $\phi /\pi$. In figures (a) and (b), $\hbar c =1$ and $-1$ 
respectively.}

\end{document}